\documentclass[%
 reprint,
 superscriptaddress,
 amsmath,amssymb,
 aps,
 pra,
]{revtex4-1}

\usepackage{graphicx}
\usepackage{dcolumn}
\usepackage{bm}

\newcommand{\eref}[1]{Eq.~(\ref{#1})}
\newcommand{\Eref}[1]{Equation~(\ref{#1})}
\newcommand{\fref}[1]{Fig.~\ref{#1}}
\newcommand{\Fref}[1]{Figure~\ref{#1}}

\begin{document}

\title{Power-Law Intensity Distribution of $\gamma$-decay Cascades\\
--- Nuclear Structure as a Scale-Free Random Network ---
} 

\author{Keisuke Fujii}
 \email{fujii@me.kyoto-u.ac.jp}
 \affiliation{%
Department of Mechanical Engineering and Science,
Graduate School of Engineering, Kyoto University
Kyoto 615-8540, Japan}
\affiliation{%
Max-Planck-Institut f\"ur Kernphysik, Saupfercheckweg 1, 69117 Heidelberg, Germany}

\author{Julian C. Berengut}%
\affiliation{%
School of Physics, University of New South Wales, New South Wales 2052, Australia
}%
\affiliation{%
Max-Planck-Institut f\"ur Kernphysik, Saupfercheckweg 1, 69117 Heidelberg, Germany}

\date{\today}

\begin{abstract}
By modeling the transition paths of the nuclear $\gamma$-decay cascade using a scale-free random network, we uncover a universal power-law distribution of $\gamma$-ray intensity $\rho_I(I) \propto I^{-2}$, with $I$ the $\gamma$-ray intensity of each transition. This property is consistently observed for all datasets with a sufficient number of $\gamma$-ray intensity entries in the National Nuclear Data Center database, regardless of the reaction type or nuclei involved. In addition, we perform numerical simulations which support the model's predictions of level population density.
\end{abstract}

\maketitle


The line intensity distribution for many-electron atoms (one of the most well-studied fermionic many-body systems) in plasmas has been reported to exhibit a power-law dependence \cite{Learner1982, Bauche-Arnoult1997, Bauche2015, Pain2013},
\begin{equation}
    \rho_I(I) \propto I^{-b},
\end{equation}
where $\rho_I(I)$ is the number of emission lines with intensity $I$ and $b$ is the index of the power-law. 
In our earlier work \cite{Fujii2020}, we showed that the index can be written as $b = 2 T_\mathrm{ex} / T + 1$, where $T_\mathrm{ex}$ is the excitation temperature in the plasma (in the energy scale), and $T$ is an atom-dependent constant related to the level density of the atom \footnote{ We use $\epsilon_0$ in Ref. \cite{Fujii2020} instead of $T$}.
This relation was derived based on two general principles: the stochastic property of the transition rates; and the exponential energy dependence of the level density of fermionic many-body systems. Therefore, we may expect other systems with the same properties, such as heavy nuclei, to exhibit a similar intensity power-law.
However, the relationship to other power-laws observed in many diverse fields of science has not yet been elucidated~\cite{Yakovenko2009,Simkin2011,Kawamura2012,Markovic2014,Munoz2018}.

Recently, it was shown that a random-walk on a general network structure typically exhibits a power-law distribution~\cite{Perkins2014,Corominas-Murtra2015, Corominas-Murtra2016}.
For example, Corominas-Murtra \emph{et al.} showed that the probability of a node being visited during a random-walk on an acyclic network follows a power-law ~\cite{Corominas-Murtra2015, Corominas-Murtra2016}.
In this Letter, we discuss the connection between quantum many-body systems and a scale-free random network such as this.
Our approach is philosophically similar to Wigner's modeling of the Hamiltonian of a quantum many-body system using a random matrix (see, e.g.~\cite{MadanLalMehta2004}): here instead, we model the optical transition paths using a scale-free random network, where the levels of the quantum system correspond to the nodes of the network, while the transitions correspond to the edges.

\begin{figure*}
    \centering
    \includegraphics[width=17cm]{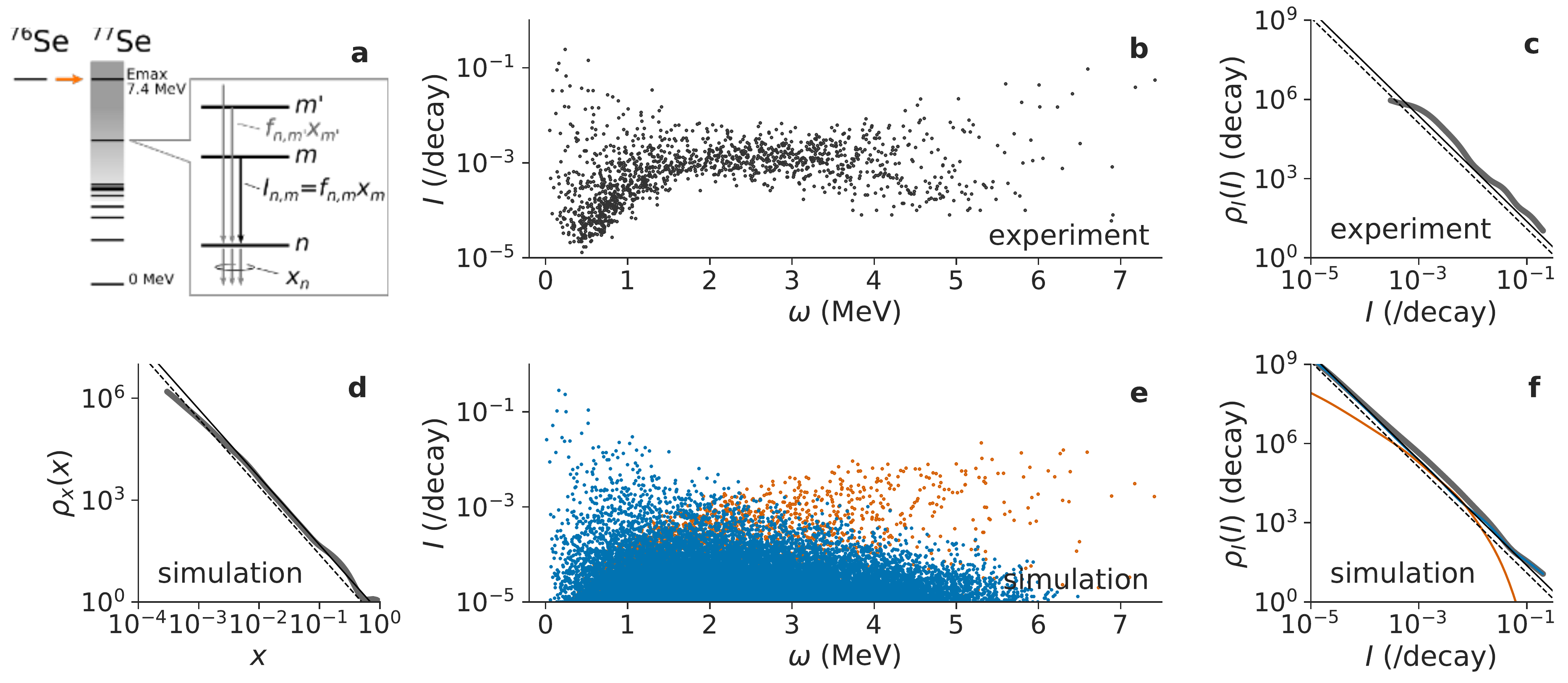}
    \caption{
      (a) Schematic illustration of the $\gamma$-decay cascade after the thermal neutron capture of $^{76}\mathrm{Se}$.
      (b) and (c) experimentally observed distributions of $\gamma$-rays resulting from this process.
      (d) -- (f) distributions simulated by \texttt{RAINIER}.
      (b) and (e) joint distributions of the $\gamma$-ray intensity and energy, respectively.
      (c) and (f) intensity distributions of the $\gamma$-ray.
      (d) Simulated distribution of $x$, which is the probability of experiencing a particular level during the cascade.
      Solid and dotted straight lines in (c), (d), and (f) show the theoretical results with $\eta = 12 / 25$ and $\eta = 1 / 4$, respectively.
      In (e) and (f), the contribution from the resonance state is shown in a different color.
    }
    \label{fig:spectra}
\end{figure*}

In particular, we discuss $\gamma$-decay cascades of heavy nuclei.
After a nuclear reaction, such as thermal neutron capture, an excited nucleus at a particular excited level, of which the atomic number has been incremented, is generated.
The excited nucleus decays to a lower level by emitting a $\gamma$-ray photon until it reaches the ground state (see \fref{fig:spectra} (a) later).
A nucleus cannot be excited to any upper levels again, and thus, as the cascade proceeds, the number of levels that can decay is reduced.
This is very similar to the sample-state-reducing (SSR) process discussed in Refs.~\cite{Corominas-Murtra2015, Corominas-Murtra2016}.
We show that such a process on a scale-free random acyclic network exhibits a universal power-law $\rho_I(I) \propto I^{-2}$,
where $I$ is the probability of passing a particular edge during the cascade and $\rho_I$ is the number density of edges with the probability $I$.
In the $\gamma$-decay cascade, $I$ corresponds to the intensities of the $\gamma$-decay transitions.
We demonstrate below that all large datasets comprising $\gamma$-ray intensities that are stored in the ENSDF database \cite{ENSDF} comply with this intensity power-law regardless of the nuclear structure and reaction type.

We begin by considering a cascade process on an acyclic network --- a directed network with no cycles --- consisting of $N$ nodes \cite{Karrer2009}.
Acyclic networks have a natural ordering and the indices $1, \cdots, N$ are assigned to each node according to this ordering.
The cascade process starts from the $N$-th node.
Let $r_{n,m}$ be the weight of an edge from the $m$-th node to the $n$-th node, and assume that the probability of jumping from the $m$-th node to the $n$-th node in one step (the branching ratio) is determined from this weight as $f_{n,m} = r_{n,m} / r_{m}$, where $r_m = \sum_{n < m} r_{n,m}$.
Note that the discussions below can be generalized to unweighted networks by assuming binary weights $r_{n,m} \in \{0, 1\}$.
The total probability of passing the edge $n \gets m$ during the cascade can be written as
\begin{equation}
    \label{eq:intensity}
    I_{n,m} = f_{n,m} x_m,
\end{equation}
where $x_m$ is the total probability of the decay passing through the $m$-th node, which satisfies
\begin{align}
    \label{eq:population}
    x_n = \sum_{m > n}^N f_{n,m} x_m.
\end{align}
Corominas-Murtra \emph{et al.}~\cite{Corominas-Murtra2015} showed that certain forms of $r_{n,m}$ give the power-law distribution of $x$.
Here, we further expand upon sufficient conditions on $r_{n,m}$ to give rise to the power-law.

We assume that $r_{n,m}$ is sampled from a certain independent random distribution and traces the cascade process on the realized network.
Let $f(n,m) = \mathbb{E}[f_{n,m}]$ be the expectation value of $f_{n,m}$. 
We assume the network is large, i.e. $N \gg 1$.
Based on this assumption, we approximate \eref{eq:population} by an integration,
\begin{align}
    \label{eq:integration_equation}
    x(n) \approx \int_n^N f(n, m) x(m) dm,
\end{align}
where $x(n) = \mathbb{E}[x_n]$.

Let us consider a class of $f(n,m)$ having the following form,
\begin{equation}
    \label{eq:sufficient_condition}
    f(n,m) = \frac{1}{m}\;g\left(\frac{n}{m}\right)
\end{equation} 
with any non-negative function $g$ satisfying $\int_0^1 g(t) dt = 1$.
This form represents the self-similarity of the edge weight distribution, and thus the network is considered as scale-free.
Then, $x(n) = \eta\, n^{-1}$ is the solution of \eref{eq:integration_equation} with a normalization constant $\eta$.
This result includes the particular case $r_{n,m} \propto n^\alpha$, which is reported to have the solution $x(n) \propto n^{-1}$~\cite{Corominas-Murtra2016}.
After the normalization, this case is reduced to $f(n,m) = \frac{\alpha + 1}{m} \left(\frac{n}{m}\right)^\alpha$, which is one realization of \eref{eq:sufficient_condition}.
On the other hand, the reported counter example $r_{n,m} \propto e^{-\beta n}$ \cite{Corominas-Murtra2016} does not satisfy \eref{eq:sufficient_condition}.

The constant $\eta$ can be evaluated by considering the total probability of the cascade proceeding from some node $m > n$ to a node $n' < n$ (i.e. the total probability flux past $n \ll N$):
\begin{align}
    1 = \int_0^n dn' \int_n^N \frac{1}{m}g\left(\frac{n'}{m}\right) x(m) dm,
\end{align}
which gives
\begin{align}
    \label{eq:normalization}
    \frac{1}{\eta} = -\int_0^1 \log t\; g(t) dt.
\end{align}
Physically, $\eta$ can be understood as a measure of the preference for connecting to a closer node of the network.
In the absence of any preference --- for example, in the case $g(t) = 1$ --- then we obtain $\eta = 1$.
If the connection between closer nodes is more probable ($g(t)$ is larger at $t \approx 1$ than at $t \approx 0$), then the integrand becomes smaller and $\eta > 1$.

At times it may be convenient to consider the density distribution of $x$, as the explicit node ordering is not always clear.
From $x(n) = \eta\, n^{-1}$, the number of nodes in the range $x \sim x + dx$ can be derived as
\begin{equation}
    \label{eq:node_distribution}
    \rho_x(x) dx = 1 dn = \eta x^{-2} dx,
\end{equation}
which is another representation of Zipf's law with the index 1.

Let us consider the joint distribution of $I$ and $t = \frac{n}{m}$, which is the number of edges within $I \sim I + dI$ and $t \sim t + dt$.
This can be computed from the unit density of $n$ and $m$, and the technique of changing random variables,
\begin{align}
    \rho_{I, t}(I, t) dI dt = 1 dn dm = g(t) \frac{\eta}{2}I^{-2} dI dt.
\end{align}
By integrating $\rho_{I, t}$ over $t$, we arrive at the power-law distribution of $I$,
\begin{align}
    \label{eq:final}
    \rho_I(I) = \int_0^1 \rho_{I, t}(I, t) dt = \frac{\eta}{2} I^{-2}.
\end{align}
In this derivation, we neglect the stochastic property of the edge weight and only consider the mean value of $f$.
However, the random fluctuation of $f$ does not change the intensity distribution, as we show in Supplemental Material.

Now, let us return to our particular system of interest, the $\gamma$-decay cascade of nuclei.
\Fref{fig:spectra}~(a) shows a schematic illustration of the $\gamma$-decay cascade of $^{77}\mathrm{Se}$ after a nuclear reaction, specifically, the thermal neutron capture of $^{76}\mathrm{Se}$.
In the nuclear case, $r_{n,m}$ is the physical transition rate from the $m$-th to the $n$-th level, and thus $f_{n,m}$ is the branching ratio.
$x_m$ is the total probability that the cascade passes through the $m$-th level, and hence
$I_{n,m} = f_{n, m} x_m$ is the total probability of the transition $n \gets m$ and is measured as the $\gamma$-ray intensity for each transition.
The $\gamma$-ray spectrum has been measured for various kinds of reactions.
In \fref{fig:spectra}~(b), we show the intensity and energy distribution of the $\gamma$-rays for the thermal neutron capture of $^{76}\mathrm{Se}$, which are taken from the ENSDF database \cite{ENSDF}.

The distribution of the transition rates has been discussed for a long time, as this is a key parameter to explain the observed abundance of elements \cite{Tveten2018}.
To consider this property of the $\gamma$-cascade spectrum, following an existing approach \cite{Hagiwara2019}, let us start from the level density of a nucleus $\rho_E(E)$, i.e., the number of levels with a given excited energy $E$ per unit energy.
In general, $\rho_E(E)$ is nearly exponentially dependent on $E$.
A simple yet well-accepted approximation thereof is the constant temperature model \cite{VonEgidy1988},
\begin{equation}
    \label{eq:rhoE}
    \rho_E(E) = \rho_0 e^{E/T}
\end{equation}
with $\rho_0 = e^{-E_0/T} / T$,
where $T$ parameterizes the level-density growth rate with excited energy (referred to as the \textit{temperature} \cite{VonEgidy1988,Egidy2005}) and $E_0$ is the energy backshift.
Here, we assume that $T$ is constant over the entire energy range, and independent of quantum numbers, spin, and isospin.
By integrating \eref{eq:rhoE} from the ground state, the level index $n$ has the following relation to the level energy $E_n$: 
\begin{equation}
    E_n = T\log\left(
        \frac{n + \rho_0 T}{\rho_0 T}
    \right).
\end{equation}
Note that the use of a different level density model, such as the back-shifted Fermi gas model, does not change the above relation, since all of them behave similarly in the energy range smaller than the resonance level.

The transition rates are expected to fluctuate according to the level pairs. On the other hand, the average rates are considered to depend on the transition energy but to be independent of the state \cite{Brink1955, Axel1962, Guttormsen2016, Brody1981,Mitchell2010,Weidenmuller2009}.
In particular, the transition rate from state $m$ to state $n$ is often written as $2 \pi \omega_{n,m}^3 \Gamma(\omega_{n,m})$, where $\omega_{n,m} = E_m - E_n$ is the energy difference and $\Gamma(\omega_{n,m})$ is the so-called gamma strength function.
(Although we implicitly assume the dominance of electric or magnetic dipole transitions, the effect of quadrupole transitions can be absorbed by $\Gamma$.)
As the energy dependence of the gamma strength function is not large, we may approximate the transition rate by taking its leading order: 
\begin{equation}
    r(n,m) \propto (E_m - E_n)^3 = \left( -T\log\left(
        \frac{n + \rho_0 T}{m + \rho_0 T}\right) \right)^3.
\end{equation}
This function has a $\left[-\log\left(\frac{n}{m}\right)\right]^3$ dependence with a large $m \gg \rho_0 T$, whereas it is proportional to $\left(1 - \frac{n}{m}\right)^3$ with small $m \ll \rho_0 T$. Both of these satisfy \eref{eq:sufficient_condition}, and so the $\gamma$-cascade spectrum is anticipated to very closely follow the power-law.
Using \eref{eq:normalization}, we obtain $\eta = \frac{1}{4}$ for the case of large values of $m$, and $\eta = \frac{12}{25}$ for the case of small values of $m$.
 
Note that the detailed energy dependence of $\Gamma$, which has been discussed frequently, slightly changes the normalization value $\eta$ but not the power-law index.
We have confirmed that reasonable profiles, such as the Lorentzian dependence of $\Gamma$ \cite{Axel1962,Berman1975}, the pygmy dipole resonances \cite{Voinov2001,Krticka2004,Simon2016}, and the low-energy enhancement \cite{Voinov2004,Karampagia2017}, do not change the value of $\eta$ significantly.
Also note that the transition selection rules do not change the intensity distribution provided that the distributions of the total angular momentum number $J$ and the parity are uniform over the energy range, since this effect can be absorbed into fluctuations of $f$.

\begin{figure}
    \centering
    \includegraphics[width=7cm]{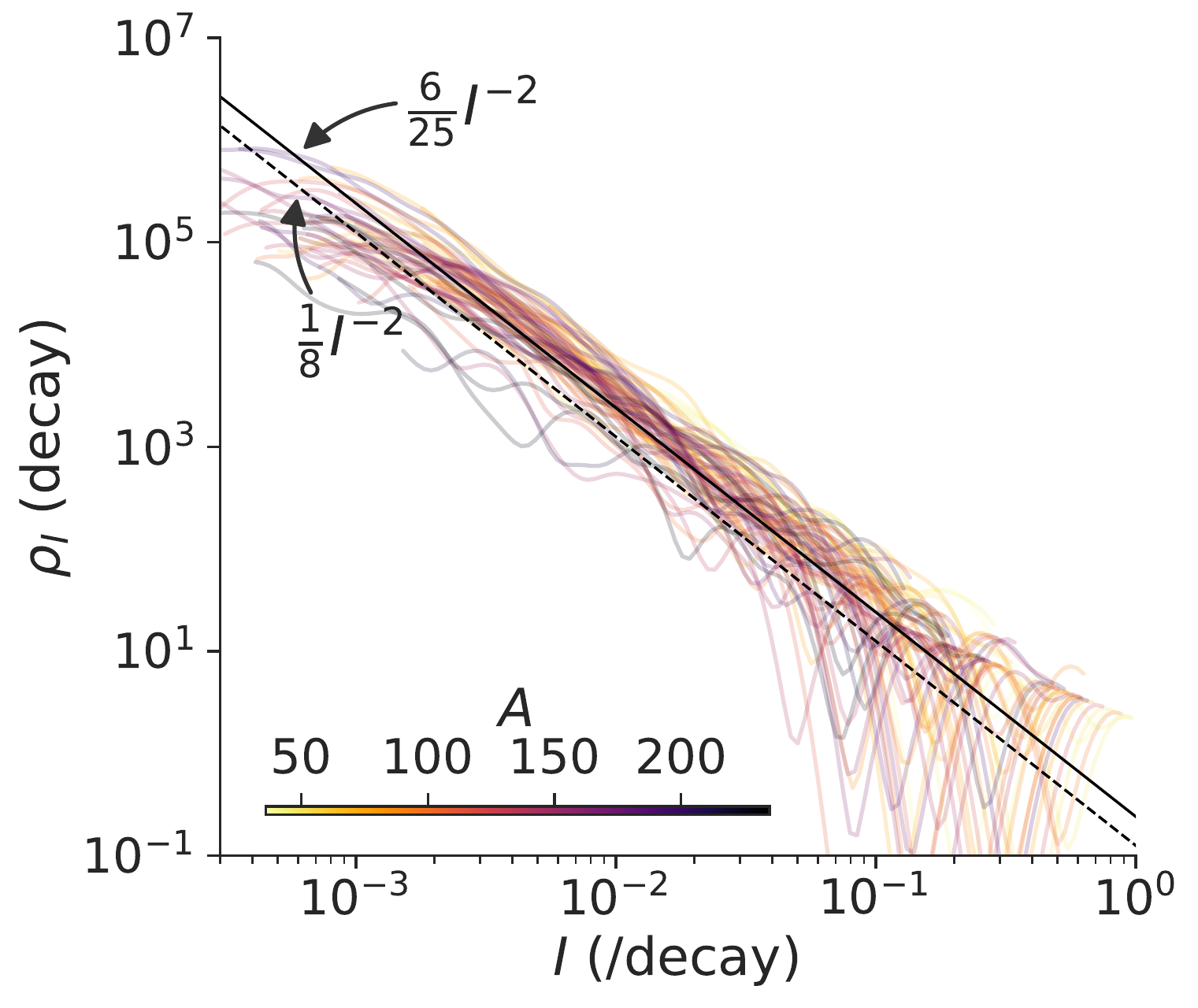}
    \caption{
        Intensity histogram of $\gamma$-cascade spectra for all the reactions with more than 200 $\gamma$-ray intensity entries in the ENSDF database \cite{ENSDF}.
        The color of each line indicates the mass number of the nucleus $A$.
        The solid and dashed lines indicate \eref{eq:final} with $\eta=12/25$ and $1/4$, respectively.
    }
    \label{fig:all}
\end{figure}

\Fref{fig:spectra} (c) shows the intensity distribution of the $\gamma$-decay cascade for a thermal neutron capture of $^{76}\mathrm{Se}$, which is computed from the observed spectrum shown in \fref{fig:spectra} (b) by the kernel density estimation method.
The diagonal solid and dashed lines show the power-law $\frac{\eta}{2}I^{-2}$, with $\eta = \frac{12}{25}$ and $\eta=\frac{1}{4}$, respectively.
This experimental distribution is consistent with the theory.

In our subsequent investigations, we compared the results of the above model with those of a numerical simulation.
\texttt{RAINIER} is a simulation tool for distributions of excited nuclear states and cascade fluctuations \cite{Kirsch2018}.
This tool adopts more exact nuclear properties, for example, the back-shifted Fermi gas model is used for the level density, taking into account parity and angular momentum distributions, certain known low-lying energy levels, and the generalized Lorentzian form for the gamma strength function. 
The level density parameters and gamma strength function parameters used in our simulation were taken from Refs.~\cite{Egidy2005} and~\cite{KIM2007}, respectively.

\Fref{fig:spectra} (e) shows the simulated intensity-energy distribution of the $\gamma$-ray lines.
This is consistent with the experiment (\fref{fig:spectra} (b)), but the distribution is extended to the much weaker intensity side than in the actual observation.
\Fref{fig:spectra} (f) shows the simulated intensity distribution, which is very close to that predicted by \eref{eq:final} over more than four orders of magnitude.
In the simulation, we can also compute the distribution of $x$ directly.
\Fref{fig:spectra} (d) shows $\rho_x(x)$, which we find is also consistent with the theory $\eta x^{-2}$, shown by diagonal lines.

\begin{figure}
    \centering
    \includegraphics[width=8.5cm]{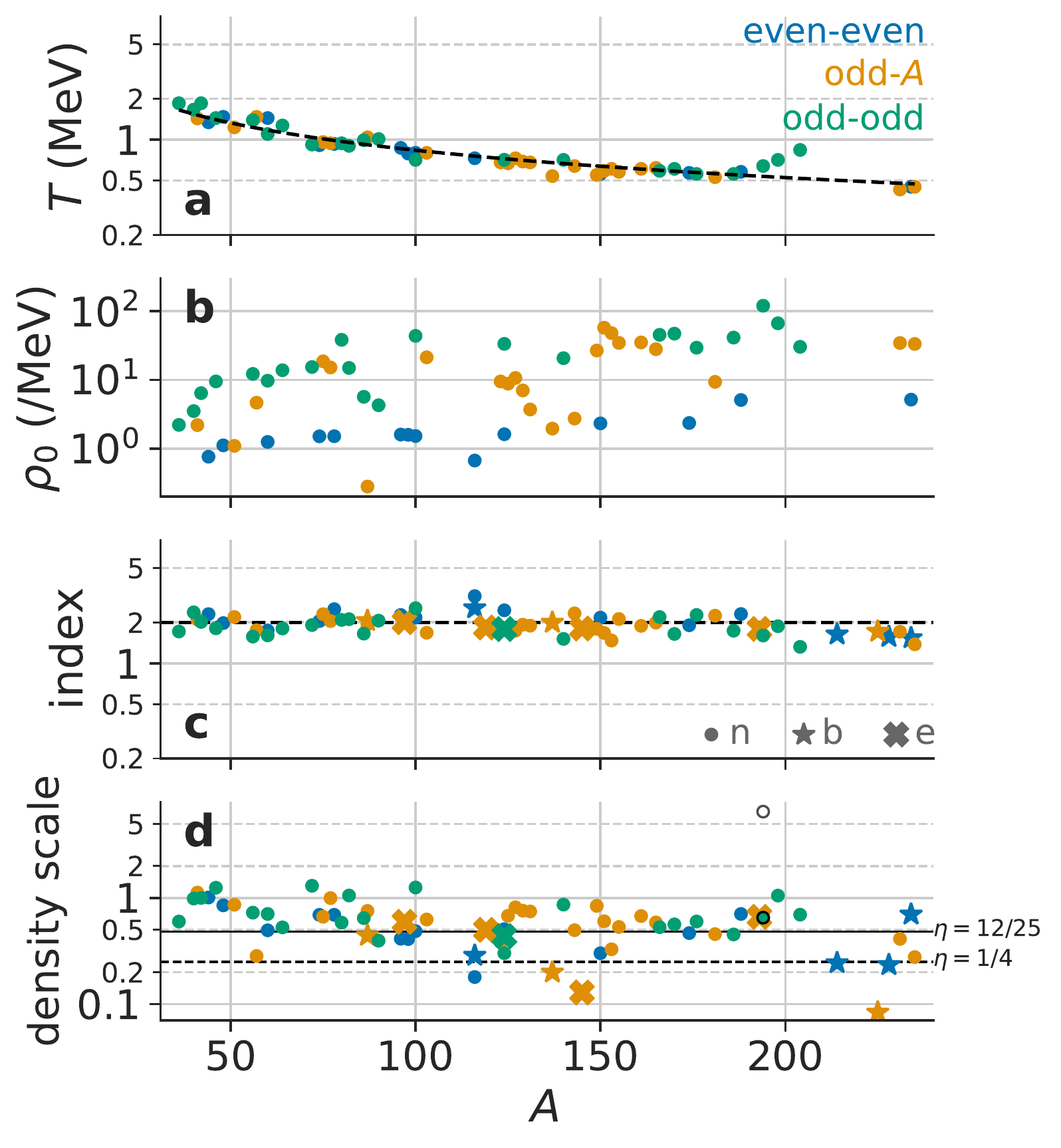}
    \caption{
        Fitted results for the $\gamma$-ray intensity distribution.
        (a) Temperature parameters $T$ and (b) level density scale $\rho_0$ for each nucleus \cite{Egidy2005}.
        (c) Index and (d) density scale of experimental $\gamma$-ray intensity distribution.
        The color of each marker indicates the type of nucleus and its shape (circle, star, or cross) indicates the reaction type (neutron capture, $\beta$-cascade, or electron capture, respectively).
        Although $T$ varies by a factor of 5 and $\rho_0$ varies by more than $10^3$, the index and density scale of the $\gamma$-ray intensity distribution are very close to the theoretical values (2 for the index and approximately 1/4 -- 12/25 for the density scale).
        The open circle in (d) shows the density scale for $^{193}\mathrm{Ir}$ from ENSDF, the intensities of which we found to be incorrect in the database by a factor of 10.
        The result for the corrected spectrum is shown by a filled circle.
    }
    \label{fig:index}
\end{figure}

To confirm the universality of this distribution, we generated intensity distributions for all $\gamma$-decay cascade spectra with more than 200 $\gamma$-intensity entries in the the ENSDF database~\cite{ENSDF}.
\Fref{fig:all} shows the intensity distributions of these 69 experimentally observed $\gamma$-cascade spectra.
The level density parameters $T$ and $\rho_0$ and the symmetry of the nuclei analyzed in this work are also shown in \fref{fig:index} (a) and (b).
As shown in \fref{fig:all}, all the distributions are concentrated on a single straight line in the double-logarithmic graph.
\Eref{eq:final} with $\eta=1/4$ and $\eta=12/25$ is plotted using black lines. 
The consistency is very clear over several orders of magnitude.

We estimated the distribution variation by fitting them with $\rho_I(I) = \frac{a}{2} I^{-b}$, where $b$ and $a$ are adjustable parameters to be estimated from the distribution, index, and density scale, respectively.
The results are shown in \fref{fig:index} (c) and (d).
Although $T$ depends on the negative power of mass number $A$ ($\approx A^{-2/3}$ \cite{Egidy2005}), which changes over a factor 5 in this range, and $\rho_0$ varies over three orders of magnitude depending on the nuclear structure, the power-law index is concentrated in the small range $b = 2.0 \pm 0.3$, which our theory predicts to be $2$. The density scale is distributed over the range $a = 0.54 \pm 0.28$ per cascade, which is predicted to be approximately in the range $1/4 \sim 12 / 25$.
The consistency with \eref{eq:final} indicates that that the scale-free random network model robustly reflects the nature of the nuclear structure.
Note that the finite-size effect of the network affects the intensity distribution to a certain extent but at the same time avoids the divergence of the total $\gamma$-ray photon number during the cascade;
see Supplemental Material for details.

Because \eref{eq:final} is a universal distribution and does not depend on the nuclear structure or reaction type, it can be used to roughly calibrate the $\gamma$-ray intensity.
Indeed, our work revealed that the intensities of $^{193}\mathrm{Ir}$ stored in ENSDF are incorrect in that they differ from those reported in the original work \cite{Balodis1998} by a factor of 10. 
This outlying behavior of this dataset is very obvious (Fig.~\ref{fig:index}(d)).

In this study, we have considered the pure decay process, in which the system can only transit to the lower levels without any excitation processes to the upper levels.
By contrast, many-electron atoms in plasmas (for which the intensity power-law was originally observed) can experience excitation.
Corominas-Murtra \emph{et al.}~\cite{Corominas-Murtra2015, Corominas-Murtra2016} considered a similar process in which they mixed the decay process (consisting of only transits to lower levels) and the bidirectional jump process (in which excitation to higher levels could also occur in the system).
They referred to this process as a noisy SSR process and showed that, by mixing these two processes, the exponent of the power-law varies in the range 1 -- 2.
This is consistent with our explanation \cite{Fujii2020}, where the exponent changes depending on the plasma parameters.
This strongly suggests a direct connection between the atomic systems and the noisy SSR process.
However, this relation would have to be further investigated in future.

In summary, we pointed out that the nuclear structure can be modeled using a scale-free random network.
We showed that a random-walk on a scale-free random acyclic network exhibits a power-law distribution for the number of edges with a given passing probability, which corresponds to the intensity of the $\gamma$-ray for the $\gamma$-cascade.
All experimental $\gamma$-ray intensities stored in the ENSDF database adhere to this power-law.

\begin{acknowledgments}
    This work was partly supported by JSPS KAKENHI Grant Number 19K14680, a grant from the Joint Research by the National Institutes of Natural Sciences (NINS) (NINS program no. 01111905), and partly by the Max-Planck Society for the Advancement of Science. JCB is supported by the Alexander von Humboldt Foundation. We thank Jos\'e Crespo L\'opez-Urrutia,  Wenjia Huang, and Hans Arwed Weidenm\"uller, for their useful discussions.
\end{acknowledgments}

\bibliography{refs}

\pagebreak
\widetext
\begin{center}
\textbf{\large Power-Law Intensity Distribution of $\gamma$-decay Cascades\\
--- Nuclear Structure as a Scale-Free Random Network ---}
\end{center}
\setcounter{equation}{0}
\setcounter{figure}{0}
\setcounter{table}{0}
\makeatletter
\newcommand{\EqIntegration}{4}
\newcommand{\FigSpectra}{1}

\renewcommand{\theequation}{S\arabic{equation}}
\renewcommand{\thefigure}{S\arabic{figure}}

\section{Stochastic Effect on the Network Structure}

In the main text, we neglect the stochastic effect on the network weights, i.e., the decay rates.
In this section, we show that introducing random variation of the rates does not affect the intensity distribution.

Let us assume that the distribution of weight can be written by
\begin{equation}
    f_{n,m} = r \frac{1}{m}g\left(\frac{n}{m}\right),
\end{equation}
where $r$ is an independently distributed random variable according to a certain probability distribution $p_r(r)$, which satisfies $\mathbb{E}[r] = \int r p_r(r) dr = 1$.
In nuclear physics, the Porer-Thomas distribution has been frequently assumed for $r$.
Furthermore, $r$ may additionally include the selection rules for transitions.

Even with this random variation in $f$, Eq. (\EqIntegration) may not change significantly if $N$ is large, since the integration over many states effectively averages out the randomness.
Therefore, $x(n) = \eta\,n^{-1}$ still holds and it is sufficient to consider the variation in the intensity, $I_{n,m} = r \frac{\eta}{m}g\left(\frac{n}{m}\right) n^{-1}$.
Let us define a new variable $\overline{I}_{n,m} = \frac{\eta}{m}g\left(\frac{n}{m}\right) n^{-1}$, which is the intensity without the random fluctuation in $f$ and therefore should follow the power-law distribution $\rho_{\overline{I}}(\overline{I}) = \frac{\eta}{2}\overline{I}^{-2}$ according to the discussion in the main text.
We obtain the distribution of $I = r \overline{I}$, which is the product of the two independent variables, $\overline{I}$ and $r$, as follows,
\begin{equation}
   \rho_I(I) = 
   \int_0^\infty \frac{1}{r} \rho_{\overline{I}}\left(\frac{I}{r}\right) p_r(r) dr = \frac{\eta}{2} I^{-2} \int_0^\infty r p_r(r) dr = \frac{\eta}{2} I^{-2}.
\end{equation}
This is exactly the same power-law distribution without the random variation in $f$, both in its index and scale.

\section{Effect of $\gamma$-decays from the resonance levels}

During the derivation of the $I^{-2}$-law in the main text, we assume the large network limit, $N \gg 1$.
However, $\gamma$-decay cascades start from the resonance levels at $N$ populated by the nuclear reaction, which have finite excited energy.
This implies that the probability for the nucleus to experience these resonance levels should be much higher than $\eta N^{-1}$ and thus the $\gamma$-rays from these levels should be also larger than those estimated from $N$.
In this section, we discuss the effect of the resonance levels in the intensity distribution.

Let us assume that a certain nuclear reaction populates $k$ levels directly and $\gamma$-decay cascades start from these levels.
For the sake of the simplicity, the probability of experiencing one of the resonance levels is $1/k$.
For the case of thermal neutron capture, $k$ should be in the order of unity.
There are $k \times N$ possible transitions including 0-intensity lines which are disallowed by the selection rules.
Their intensities without the fluctuation discussed in the previous section $\overline{I}$ are written by
\begin{equation}
    \overline{I}_{n,N} = f(n,N)\; x_N = \frac{1}{kN}\; g\left(\frac{n}{N}\right).
\end{equation}
Thus, the distribution of the non-fluctuated intensity from the resonance states $\rho_{\overline{I}}^\mathrm{res}$ is
\begin{equation}
    \rho_{\overline{I}}^\mathrm{res}(\overline{I})d\overline{I} = k dn 
    = \frac{k^2N^2}{g'\left(g^{-1}(kN \overline{I})\right)} d\overline{I},
\end{equation}
where $g^{-1}$ is the inverse function of $g$, i.e., $\frac{n}{N} = g^{-1}(kN \overline{I})$ and $g'(t)$ is the derivative of $g(t)$.

In order to obtain the $\rho_{\overline{I}}^\mathrm{res}$, we may need an actual form of $g$.
According to the main text, $g$ can be written as $g(t)\approx \frac{1}{6}[-\log (t)]^3$ for highly excited states, where the factor $\frac{1}{6}$ comes from the normalization condition of $g$.
By substituting this form, we obtain
\begin{equation}
    \label{sup:eq:weibull}
    \rho_{\overline{I}}^\mathrm{res}(\overline{I}) = 
    kN \cdot 3 (6 k N) \left(6kN \overline{I}\right)^{-\frac{2}{3}} 
    e^{-(6kN \overline{I})^{1/3}}
    = \mathcal{W}\left(kN\overline{I}; \frac{1}{6}, \frac{1}{3}\right),
\end{equation}
where $\mathcal{W}(z; z_0, k) = \frac{k}{z_0}\left(\frac{z}{z_0}\right)^{k-1} e^{-(z/z_0)^k}$ is the Weibull distribution of $z$ with scale parameter $z_0$ and shape parameter $k$.

\begin{figure*}
    \centering
    \includegraphics[width=13cm]{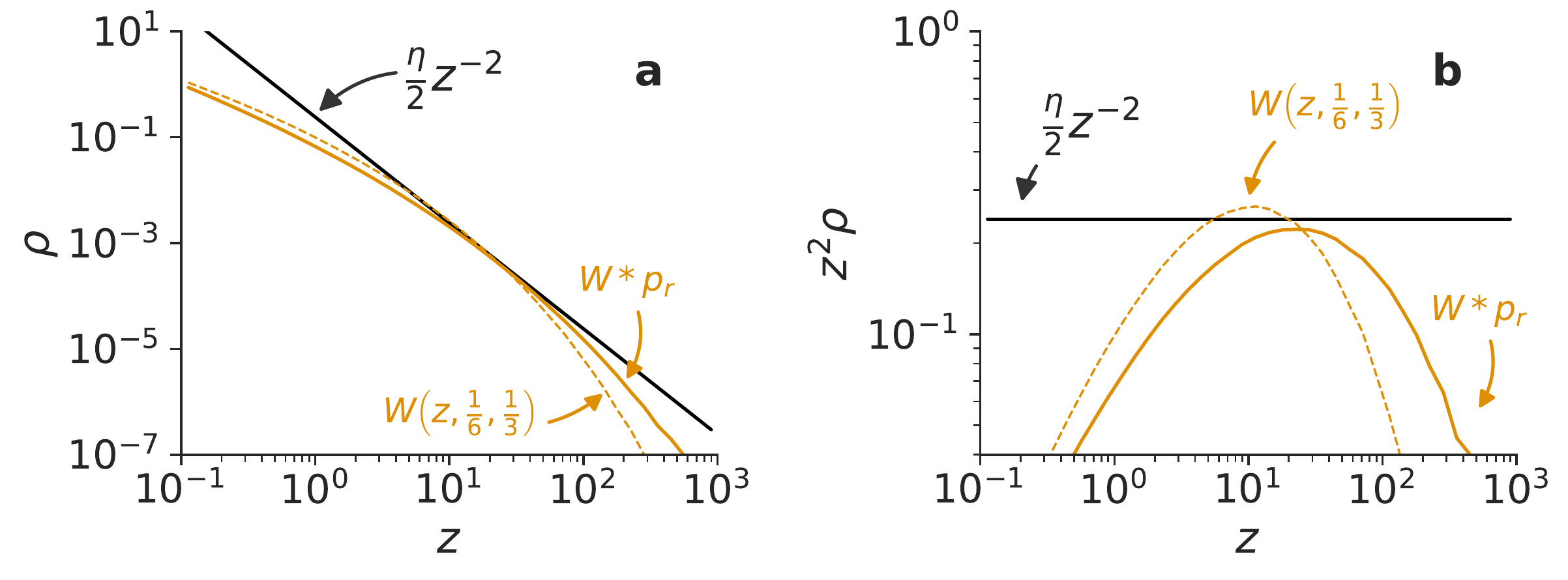}
    \caption{
      (a) Comparison between the power-law distribution with $\eta = \frac{12}{25}$ (black line) and the Weibull distribution (\eref{sup:eq:weibull}, colored dashed curve) as well as that convoluted with the Porter-Thomas distribution (colored solid curve).
      The corresponding distribution by the simulation \texttt{RAINIER} can be found in Fig.~\FigSpectra (f).
      (b) The same distributions as (a) normalized by $z^{-2}$.
      The convoluted Weibull distribution (colored solid curve) is less significant than the power-law distribution (black line).
    }
    \label{sup:fig:weibul}
\end{figure*}

The Weibull distribution with shape parameter smaller than 1 behaves similar to power-law distributions. 
Figure \ref{sup:fig:weibul}(a) shows the Weibull distribution $\mathcal{W}(z; \frac{1}{6}, \frac{1}{3})$ and the power-law distribution $\frac{6}{25}z^{-2}$  we obtained in the main text, in log-log scale.
They show similar behaviors to each other and the Weibull distribution is mostly smaller than the power-law distribution.

Except for the power-law distribution with index 1, which is discussed in the previous section, the product convolution with another random variable generally changes the distribution.
We computed the product convolution of the Weibull-distributed variable and another random variable drawn from the Porter-Thomas distribution.
The solid colored curve in Fig.~\ref{sup:fig:weibul}(a) shows the convoluted distribution, $\rho_I^\mathrm{res}$.
The convolution makes this distribution even less significant.
In order to highlight their difference, we show in Fig.~\ref{sup:fig:weibul}(b) these distributions again but scaled by $z^{-2}$.
We can conclude that the $\gamma$-ray lines from the resonance states does not change the entire intensity distribution.
This is also consistent with the simulation result by \texttt{RAINIER}, which is shown in Fig.~\FigSpectra (f) in the main text.

The finite network size (the finite resonance energy) also avoids the divergence in the number of $\gamma$-ray photons, 
$\int_{I_{min}}^1 I \rho_I(I) dI$,
which diverges if we take the limit $I_{min} \rightarrow +0$.
However, $I_{min}$ is determined by the network size $N$ and the typical value of the branching ratio $\overline{f} \approx \frac{1}{N}$, i.e., 
\begin{equation}
    I_{min} \approx \eta N^{-2}.
\end{equation}
Note that since $N$ is typically more than $10^4$, the intensity cutoff is far below the instrumental resolution and is not observed in practice.
    
\end{document}